\documentclass[universe,article,accept,pdftex,moreauthors]{Definitions/mdpi}
\firstpage{1} 
\pubvolume{1}
\issuenum{1}
\articlenumber{0}
\pubyear{2026}
\copyrightyear{2026}
\externaleditor{Firstname Lastname} 
\datereceived{28 June 2026} 
\daterevised{23 August 2026} 
\dateaccepted{25 August 2026} 
\datepublished{ }

\newcommand{\apj}{Astrophys.~J.}
\newcommand{\apjl}{Astrophys.~J.~Lett.}

\newcommand{\mnras}{Mon.~Not.~R.~Astron.~Soc.}

\newcommand{\pasp}{Publ.~Astron.~Soc.~Pac.}
\newcommand{\pasa}{Publ.~Astron.~Soc.~Aust.}

\newcommand{\nat}{Nature}

\usepackage{dcolumn}
\usepackage{bm}

\Title{Spectro-Polarimetric Properties of CHIME FRB Sources}

\Author{Dengke Zhou 
 $^{1}$, Yi Feng $^{1,2,}$*, Jiaying Xu $^{1}$, Chenyuan Xu $^{1}$ and Jianhua Fang $^{1}$}

\address{%
$^{1}$ \quad Research Center for Computational Earth and Space Science, Zhejiang Laboratory, Hangzhou 311100, China; zdk@zhejianglab.com (D.Z.); xujy@zhejianglab.com (J.X.); xuchenyuan@zhejianglab.org (C.X.); fangjh@zhejianglab.com (J.F.) 
\\
$^{2}$ \quad Institute for Astronomy, School of Physics, Zhejiang University, Hangzhou 310027, China
}

\corres{Correspondence: yifeng@zhejianglab.org}

\abstract{Fast radio bursts (FRBs) are enigmatic millisecond-duration radio transients whose polarization properties offer crucial insights into their origins and environments. In particular, low-frequency depolarization---quantified by the parameter \(\sigma_{\mathrm{RM}}\)---probes the complex magneto-ionic medium surrounding the progenitor, and has been observed across a population of repeating FRBs. We present a systematic spectro-polarimetric analysis of repeating and non-repeating FRBs using observations from the Canadian Hydrogen Intensity Mapping Experiment (CHIME). For 28 repeating FRBs, we measure \(\sigma_{\mathrm{RM}}\), expanding the known sample from 14 to 36 sources (an increase by a factor of 2.6). The kernel density estimate (KDE) of the repeating population peaks at \(1.3\ \mathrm{rad\,m^{-2}}\), with approximately 70\% of the sources showing \(\sigma_{\mathrm{RM}} \gtrsim 1\ \mathrm{rad\,m^{-2}}\), implying that most reside in complex magneto-ionic environments. For 70 non-repeating FRBs, we investigate four spectro-polarimetric models; no source exhibits significant depolarization with \(\sigma_{\mathrm{RM}} \gtrsim 5\ \mathrm{rad\,m^{-2}}\). Roughly half of the non-repeaters are consistent with a constant linear polarization fraction across frequency. We caution, however, that these results may be affected by the limited frequency coverage of CHIME. Future ultra-wideband polarimetry, spanning widely separated frequencies, will overcome current observational biases, enable precise \(\sigma_{\mathrm{RM}}\) measurements, and substantially deepen our understanding of FRB environments.}


\begin{document}

\section{Introduction}

Fast radio bursts (FRBs) are enigmatic millisecond-duration extragalactic radio transients. Since the discovery of the first FRB in 2007~\citep{2007Sci...318..777L}, the known FRB population has grown rapidly, driven by the increasing sensitivity and survey capabilities of modern radio telescopes. Repeating FRBs are FRBs that originate from sources that emit multiple bursts over time. According to current statistics compiled in the Blinkverse database\endnote{\url{https://blinkverse.zero2x.org}}~\citep{blinkverse}, more than 4000 FRBs have been publicly reported to date; repeating FRBs constitute only a small fraction ($\sim$2.5\%) of the total known population, while the majority ($\sim$97.5\%) have been detected as one-off events. Despite these advances, including precise localization and host-galaxy identification for a growing number of sources~\citep{2025Univ...11..206X}, the physical origin, emission mechanism, and environmental properties of FRBs remain largely unresolved~\citep{zhangreview2023}.

Polarization provides one of the most powerful diagnostics for understanding the physical nature and local environments of FRBs. Precise measurements of polarization can probe the magneto-ionic medium surrounding the source and provide important insights into the underlying radiation mechanism and propagation effects along the line of sight. The linear polarization position angle rotates as the radio wave propagates through magnetized plasma, and the magnitude of this effect is quantified by the rotation measure (RM), which reflects the integrated electron density and magnetic field along the line of sight. The RM of repeating FRBs may vary depending on their \mbox{environments \citep{yang2023}.} 
Observationally, a significant proportion of repeating FRBs show RM variations \citep{feng2025}. 
The RM of FRB~20121102A dropped substantially over time \citep{2018Natur.553..182M, 2021ApJ...908L..10H}. RM reversals have been observed in sources including FRB~20190520B, FRB~20190117A, and FRB~20220529A \citep{reshma23, chime_repeaterRM, feng2025, liye2026}. 
An RM jump was also observed in FRB~20220529A \citep{liye2026}. 
These observations suggest that these repeating FRBs reside in dynamic magneto-ionic environments, possibly including expanding supernova remnants, pulsar wind nebulae, and massive binary \mbox{systems \citep{2018ApJ...861..150P, 2018ApJ...868L...4M, 2021ApJ...923L..17Z, 2022MNRAS.510L..42K, 2022NatCo..13.4382W, yang2023}.}

Complex environments may leave imprints on the spectro-polarimetric properties of FRBs. The linear polarization fraction may become frequency-dependent, with the magnitude of this dependence characterized by the RM scatter $\sigma_{\mathrm{RM}}$ \citep{1966MNRAS.133...67B}. Using observations from several telescopes covering 115\,MHz to 4.6\,GHz, \citet{feng22} reported depolarization behavior in repeating FRBs: the linear polarization fraction decreases steadily toward lower frequencies, quantified by $\sigma_{\mathrm{RM}}$ values between 0.12 and 218\,$\mathrm{rad\,m^{-2}}$, attributed to multipath scattering in turbulent magneto-ionic environments. This depolarization behavior not only probes the intervening magneto-ionic medium and constrains possible progenitors and emission mechanisms, but may also point to a distinctive environmental link: 
the two FRBs with the largest $\sigma_{\mathrm{RM}}$, namely FRB 20121102A ($30.9 \pm 0.4\ \mathrm{rad\,m^{-2}}$) and FRB 20190520B ($218 \pm 10.2\ \mathrm{rad\,m^{-2}}$), are both known to be associated with persistent radio \mbox{counterparts \citep{2017Natur.541...58C,feng22,niu22}}.
The linear regression between $\log|\mathrm{RM}|$ and $\log\sigma_{\mathrm{RM}}$ for the repeating FRB sample yields a slope of $0.62 \pm 0.30$ \citep{feng22}.

Using the Australian Square Kilometre Array Pathfinder (ASKAP) sample, a search for depolarization was conducted in 17 apparently non-repeating FRBs~\citep{askap24, askap26}. Only FRB~20230526A shows clear frequency-dependent depolarization, with its linear polarization fraction dropping from $\sim$60\% at high frequencies (1440~MHz) to $\sim$20\% at low frequencies (1110~MHz). For the other non-repeating bursts, upper limits on $\sigma_{\rm RM}$ were derived under two scenarios: assuming 100\% intrinsic linear polarization (Burn model, denoted as $\sigma_{\rm RM}$) and allowing the intrinsic polarization to be free (modified Burn model, denoted as $\sigma_{\rm RM}'$, which yields more conservative upper limits)~\citep{askap24, askap26}. When these upper limits are compared with the $\log|\mathrm{RM}|$--$\log \sigma_{\rm RM}$ relation obtained by \citet{feng22}, they are found to be consistent only if the actual $\sigma_{\rm RM}$ values are roughly two orders of magnitude smaller than $\sigma_{\rm RM}'$~\citep{askap24, askap26}.

In this paper, we use the published polarimetric data of repeating and non-repeating FRBs from the Canadian Hydrogen Intensity Mapping Experiment (CHIME)~\citep{2024ApJ...968...50P, 2025ApJ...982..154N} to measure $\sigma_{\mathrm{RM}}$ for 28 repeating FRBs and to compare four different spectro-polarimetric models against 70 non-repeating FRBs. The paper is structured as follows: Section~\ref{sec:method} describes our method; Section~\ref{sec:res} presents the main results; and Section~\ref{sec:con} summarizes our key conclusions.

\section{Method}
\label{sec:method}
In this work, we analyze two complementary FRB samples: repeating FRBs from CHIME/FRB \citep{2025ApJ...982..154N,2023ApJ...947...83C} and non-repeating FRBs from the first CHIME/FRB baseband \linebreak  catalog \citep{2024ApJ...968...50P}.

For repeating FRBs, we adopt the polarization measurements from Ng+2025 \citep{2025ApJ...982..154N}, which include 28 repeating sources. Ng+2025 provides a comprehensive study of RM temporal variability and polarization for 28 repeaters, with a distinct emphasis from this work. Critically, Ng+2025 does not examine the $\sigma_{\rm RM}$ associated with Burn-type depolarization, nor does it perform model comparisons on the frequency dependence of $L/I$. Our $\sigma_{\rm RM}$ measurements for repeaters therefore comprise a new and complementary analysis using the published measurements from Ng+2025, rather than a reinterpretation of their results. In Ng+2025, the baseband data were localized and beamformed to each repeater's sky position, coherently de-dispersed at the structure-optimized DM, and processed through the standard CHIME/FRB polarization pipeline \citep{2021ApJ...920..138M} to obtain Stokes $I$, $Q$, $U$, $V$ dynamic spectra. The linear polarization fraction $L/I$ for each burst was then calculated by integrating $L = \sqrt{Q^2 + U^2}$ over the burst profile. 

For non-repeating FRBs, we adopt the sample from Ref.~\citep{2024ApJ...968...50P}, which consists of 89 polarized bursts from the first CHIME/FRB baseband catalog. We first retain only bursts with total signal-to-noise ratio (S/N) $\geq 6$, following the S/N $\geq 6$ threshold adopted by \citet{2025ApJ...982..154N} for linear polarization detection. This selection reduces the sample to 70 non-repeating FRBs. For each burst, we use the RM derived from QU-fitting by the CHIME/FRB polarization pipeline \citep{2021ApJ...920..138M} to apply Faraday de-rotation to the Stokes $Q$ and $U$ spectra. For non-repeating FRBs, these QU-fitting RM values are taken from the published catalog of \citet{2024ApJ...968...50P} (denoted $\rm RM_{\rm QU}$, listed in Supplementary Tables~S2 and~S3). After de-rotation, we integrate the derotated Stokes spectra over the burst envelope and divide the band into sub-bands using an adaptive binning scheme: starting from a minimum interval of 5\,MHz, each sub-band is iteratively widened in 5\,MHz steps until the total intensity S/N in that sub-band exceeds 5. This allows us to extract the linear polarization fraction $L/I$ as a function of $\lambda^2$ with finer resolution where the signal is strong while maintaining sufficient S/N in each sub-band.

For each repeating FRB, we compute $\sigma_{\rm RM}$ from the Burn model \citep{1966MNRAS.133...67B}, which describes depolarization as
\begin{equation}
P_{\rm lin}(\lambda^2) = \exp(-2 \sigma_{\rm RM}^2 \lambda^4),
\label{eq:burn}
\end{equation}
where $P_{\rm lin}$ is the linear polarization fraction at wavelength $\lambda$. For each burst, $\sigma_{\rm RM}$ is obtained by inverting Equation~(\ref{eq:burn}) using the published $L/I$ measurement. In the absence of any published estimates of systematic uncertainties in CHIME polarization data, we add a systematic uncertainty of 0.05 on $L/I$ in quadrature with the statistical errors to properly assess the uncertainties in $\sigma_{\rm RM}$. This adopted systematic uncertainty is an order of magnitude larger than the value reported for the Five-hundred-meter Aperture Spherical radio Telescope (FAST) \citep{2017ursi.conf...62D}. For FRBs with multiple bursts, we take the minimum $\sigma_{\rm RM}$ among all bursts as the representative value for the source.

For non-repeating FRBs, we compare four spectro-polarimetric models to identify the best-fit description of the spectro-polarimetric behavior: (1) the Burn model, (2) a modified Burn model allowing for non-unity high-frequency polarization, (3) a constant spectro-polarimetric model, and (4) an oscillation model \citep{2022Natur.609..685X} that captures coherent oscillations in polarization fraction. 
Motivated by Faraday conversion \citep{2022Natur.609..685X, wang2025} or instrumental effects \citep{2021ApJ...920..138M}, we introduce the oscillation model.
The oscillation model is first fit jointly to the linear and circular polarization fractions, sharing a common angular frequency $\omega$ between the two polarization components. The best-fit linear-polarization parameters are then extracted from the joint posterior samples, and the marginal likelihood for the linear polarization data alone is computed via Monte Carlo integration: for each posterior sample from the joint fit, the linear-polarization likelihood is evaluated, and the marginal likelihood is obtained as the average of these likelihood values over the samples. This enables a fair Bayesian comparison with the other three models, which are fit directly to the linear polarization data via nested sampling. All four models are compared on the same linear polarization data using Bayesian evidence, with $\Delta \log_{10} E \geq 10$ adopted as the uniform threshold for strong evidence in favor of one model over another \citep{askap24, 2008ConPh..49...71T}. In addition, we compute posterior predictive check (PPC) \emph{p}-values as a quantitative goodness-of-fit diagnostic. For bursts where the oscillation model is preferred by Bayesian evidence, we further require that the linear polarization oscillation amplitude be detected at high significance ($A/\sigma_{A} > 7$) in the joint fit, together with visual inspection of the fitted oscillation pattern, ensuring that the oscillatory structure is genuinely present rather than arising from noise fluctuations.

\begin{enumerate}
    \item {Oscillation 
 model}:\vspace{-4pt}
    \begin{equation}
    P_{\rm lin}(\lambda^2) = c + k\lambda^2 + A\sin(\omega \lambda^2 + \phi_{0}),
    \label{eq:fc}
    \end{equation}
    where $A$ is the amplitude of sinusoidal variation, $\omega$ the angular frequency (shared with circular polarization), $\phi_0$ the initial phase, $k$ the linear slope, and $c$ the baseline polarization. This model captures coherent oscillations in polarization fraction induced by propagation effects in magnetized plasma. The oscillation phases of linear and circular polarizations are roughly $\pi$ radians out of phase.

    \item Constant model:
    \begin{equation}
    P_{\rm lin}(\lambda^2) = P_0,
    \label{eq:const}
    \end{equation}
    where $P_0$ is the constant linear polarization fraction, which represents wavelength-independent linear polarization, indicative of negligible Faraday depolarization within the observational band.

    \item Burn model (Equation~(\ref{eq:burn})), describing depolarization dominated by random RM variations.

    \item Modified Burn model:\vspace{-4pt}
    \begin{equation}
    P_{\rm lin}(\lambda^2) = P_0 \exp(-2 {\sigma'_{\rm RM}}^2 \lambda^4),
    \label{eq:modburn}
    \end{equation}
    where $P_0$ is the intrinsic polarization and $\sigma'_{\rm RM}$ characterizes depolarization due to RM dispersion. This model allows the intrinsic polarization fraction (at infinite frequency) to deviate from 100\%.
\end{enumerate}

The model fitting and Bayesian evidence calculations described above are performed with custom Python (v3.11.14) scripts based on the {\tt dynesty} (v3.0.0) package for nested sampling.
To better reflect the true scatter, we rescale the uncertainties on $\sigma_{\rm RM}$ and $\sigma'_{\rm RM}$ by a factor of $\sqrt{\chi^2_\nu}$ when performing the parameter estimation, where $\chi^2_\nu = \chi^2/(N-d)$ is the reduced chi-squared statistic of the best-fit model, with $N$ data points and $d$ free parameters.

\section{Results}
\label{sec:res}

\subsection{Repeating FRBs}

\begin{figure}[H]
    \includegraphics[width=0.83\linewidth]{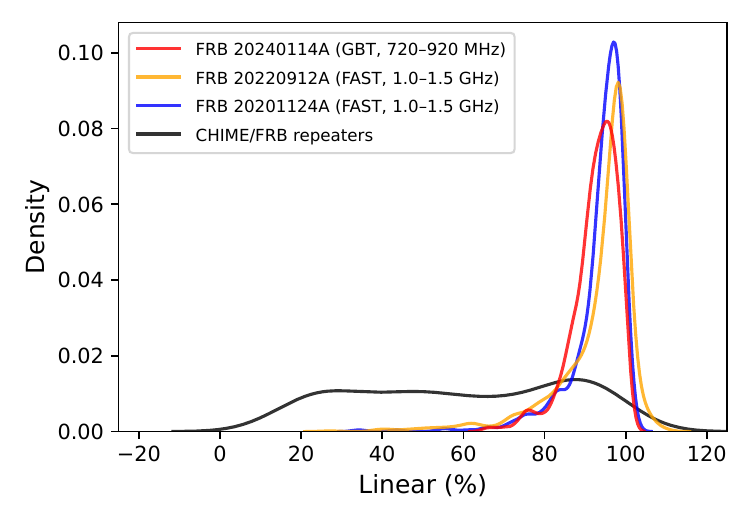}
    \caption{Linear polarization distributions of several repeating FRBs and the CHIME/FRB repeater sample. The figure presents the KDE of the linear polarization degree for three repeating FRBs (red, orange, and blue curves), together with that of the CHIME/FRB repeater sample (black curve) \citep{2025ApJ...982..154N}. The KDE tails extending beyond the physical bounds (0\% and 100\%) are artifacts of the kernel smoothing and do not represent real physical probability.} 
    \label{fig:linear_kde_repeater}
\end{figure}

Figure~\ref{fig:linear_kde_repeater} shows the kernel density estimates (KDE) of the linear polarization fraction for three repeating FRBs (FRBs~20240114A, 20220912A, and 20201124A; red, orange, and blue curves, respectively), based on measurements from Ref.~\citep{jiangraa, zhang2023, xie2024FRB20240114A} with FAST and Robert C. Byrd Green Bank Telescope (GBT), together with the KDE for our CHIME repeater sample (28 sources; black curve) \citep{2025ApJ...982..154N}.
The colored curves show a sharp peak at high linear polarization fractions ($\gtrsim$80\%--$100\%$), with median values of 95.5\%, 96.0\%, and 93.6\% for FRBs~20201124A, 20220912A, and 20240114A, respectively \citep{jiangraa, zhang2023, xie2024FRB20240114A}, all falling within this high range. By contrast, the black KDE for the CHIME sample is significantly broader and extends to much lower polarization fractions, providing clear evidence that the CHIME repeater population experiences depolarization.

\begin{figure}[H]
    \includegraphics[width=0.7\linewidth]{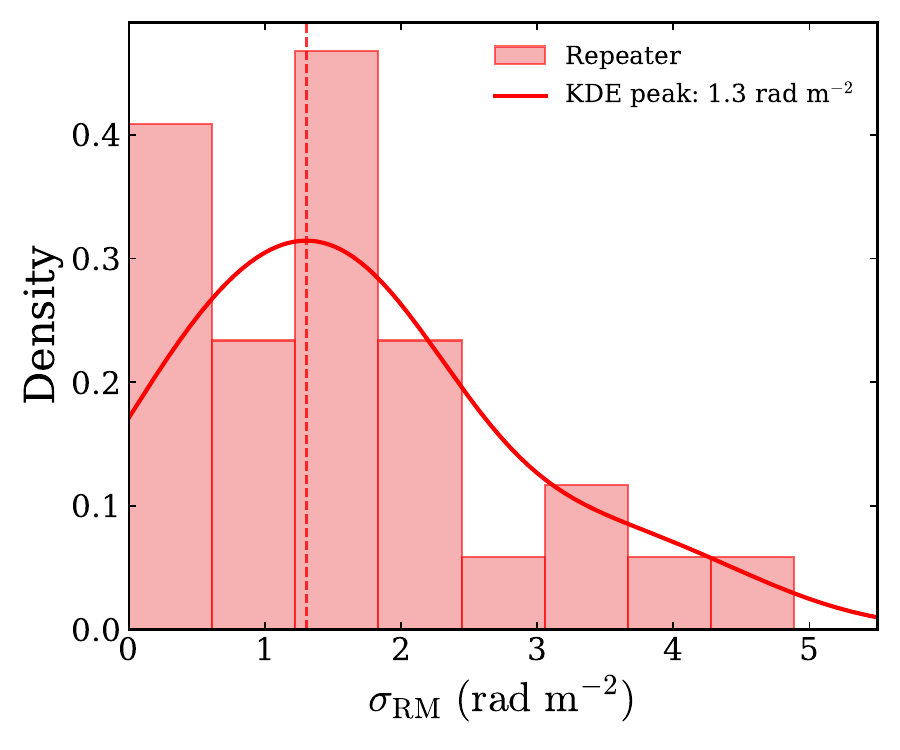}
    \caption{Distribution of \(\sigma_{\mathrm{RM}}\) for the 28 repeating FRBs. The histogram and KDE curve show the normalized density. The red dashed line marks the KDE peak at \(1.3\ \mathrm{rad\,m^{-2}}\).}
    \label{fig:sigmarm}
\end{figure}

Using the method described in Section~\ref{sec:method}, we measure \(\sigma_{\mathrm{RM}}\) for the 28 repeating FRBs. The resulting \(\sigma_{\mathrm{RM}}\) values are summarized in Supplementary Table~S1 (the corresponding RM values are from the CHIME/FRB polarization pipeline, as reported by \citet{2025ApJ...982..154N}), and the distribution of \(\sigma_{\mathrm{RM}}\) is shown in Figure~\ref{fig:sigmarm}. Previous studies reported \(\sigma_{\mathrm{RM}}\) measurements for 14 repeaters \citep{feng22, chime_repeaterRM, 2026ApJ..1000L..53P}. Our sample expands this to 36 repeaters (adding 22 new ones), representing an increase by a factor of 2.6. Approximately 70\% of the sources exhibit \(\sigma_{\mathrm{RM}} \gtrsim 1\ \mathrm{rad\,m^{-2}}\), implying that they reside in complex magneto-ionic environments.

\vspace{-10pt}

The largest \(\sigma_{\mathrm{RM}}\) measured in our sample is $\sim$$5\,\mathrm{rad\,m^{-2}}$, and we find no source with \(\sigma_{\mathrm{RM}} > 10\ \mathrm{rad\,m^{-2}}\). We interpret this absence as a consequence of selection bias. For instance, a \(\sigma_{\mathrm{RM}}\) of \(7.6\,\mathrm{rad\,m^{-2}}\) would already suppress the linear polarization fraction to $\sim$10\% at 800 MHz, the high-frequency end of the CHIME band. Given that most bursts in our sample have total S/N below 100, and that we typically require a linear polarization S/N \(> 10\) for reliable RM and polarization measurements, this 10\% level effectively defines a practical detection threshold. The absence of \(\sigma_{\mathrm{RM}}\) values larger than $\sim$$7\,\mathrm{rad\,m^{-2}}$ is therefore plausibly attributable to this same selection effect.

The KDE for the repeating population shows a peak at \(1.3\ \mathrm{rad\,m^{-2}}\). As discussed above, we caution that this peak may be subject to selection bias, since FRBs with larger \(\sigma_{\mathrm{RM}}\) may not have sufficiently high polarization S/N for reliable measurements. There is also a secondary bump at \(\sigma_{\mathrm{RM}} \lesssim 0.5\ \mathrm{rad\,m^{-2}}\), which may indicate that some repeating FRBs reside in a less complex magneto-ionic environment---a population also identified by Ref.~\citep{feng2025}. This is noteworthy given that, while a large proportion of repeating FRBs exhibit RM variations exceeding \(50\ \mathrm{rad\,m^{-2}}\), others exhibit only marginal intrinsic RM variations. Together with previously reported sources, the two FRBs with the largest \(\sigma_{\mathrm{RM}}\) are FRB~20121102A (\(30.9 \pm 0.4\ \mathrm{rad\,m^{-2}}\)) and FRB~20190520B (\(218 \pm 10.2\ \mathrm{rad\,m^{-2}}\)) \citep{feng22}. The broad range of \(\sigma_{\mathrm{RM}}\) values suggests that repeating FRBs may be at different evolutionary stages \citep{feng22, 2022Sci...375.1227C} or may even have distinct physical origins.

\subsection{Non-Repeating FRBs}
\begin{figure}[H]
    \includegraphics[width=0.9\linewidth]{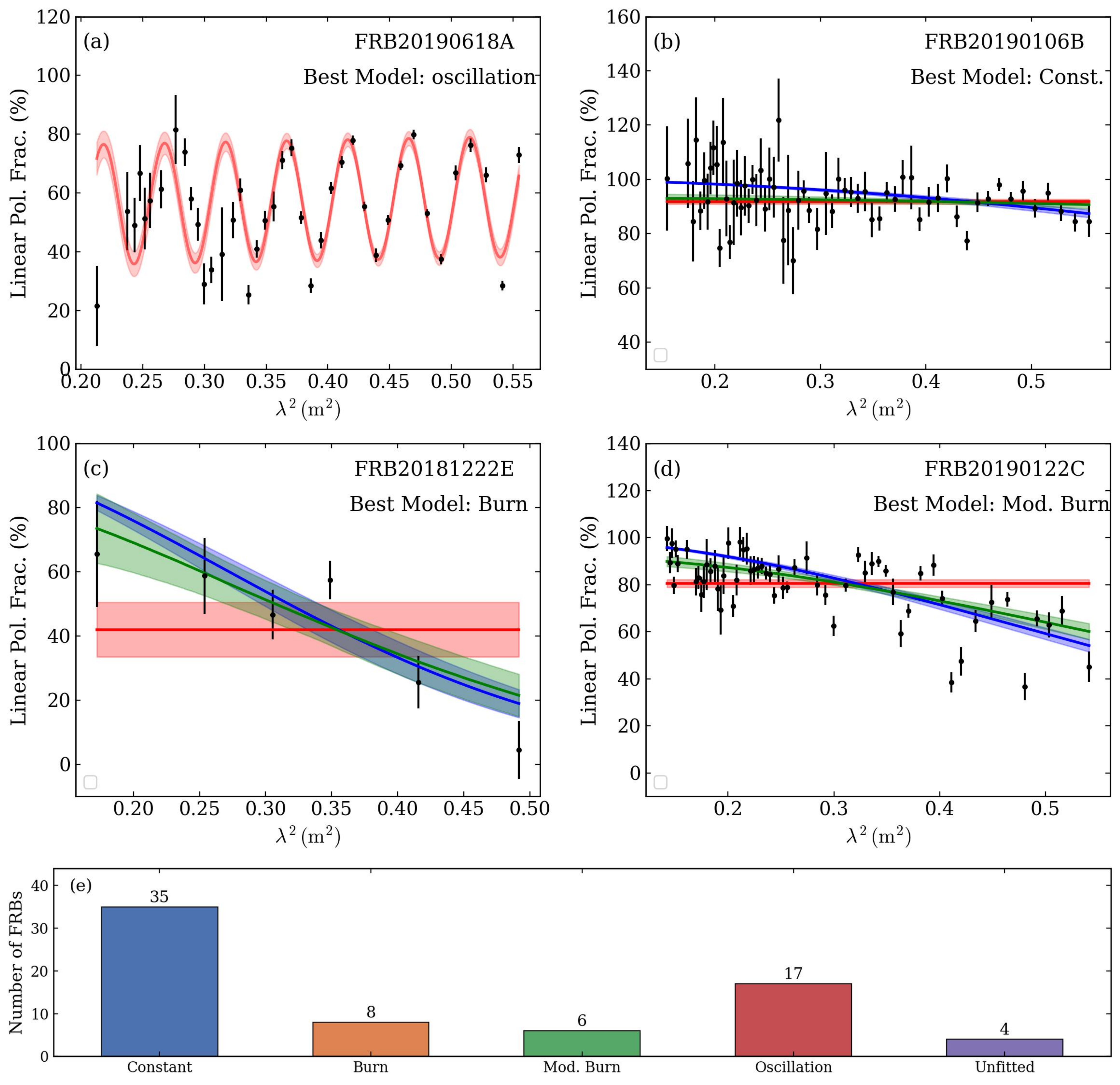}
    \caption{Fractional linear polarization vs.\ wavelength-squared for four representative non-repeating FRBs, illustrating different best-fitting models: oscillation (a), constant (b), Burn (c), and modified Burn (d). Black points with error bars show the data; shaded regions indicate $1\sigma$ confidence intervals. Oscillation fit is shown in red in panel (a); for other panels, red, blue, and green curves represent Const., Burn, and Mod.\ Burn models, respectively. The preferred model for each burst is indicated. Panel (e) shows the model distribution across the 70 non-repeating FRB sample.}
    \label{fig:FCBMB}
\end{figure}

Following the method outlined in Section~\ref{sec:method}, we perform a Bayesian model comparison among all four spectro-polarimetric models for the 70 non-repeating FRBs in our sample. For the 17 sources for which the oscillation model is strongly preferred ($\Delta\log_{10}E \geq 10$), passes the PPC diagnostic, satisfies the amplitude significance criterion ($A/\sigma_A > 7$), and is confirmed by visual inspection, we list the best-fit parameters in Supplementary Table~S2 and display the corresponding fits in Supplementary Figure~S1. Nevertheless, in the absence of any published reports of systematic instrumental effects in CHIME data, we are presently unable to distinguish between Faraday conversion, instrumental contamination, and other scenarios. Accordingly, we defer a deeper physical interpretation of the oscillating sources to future investigations. For the remaining 53 bursts, the Bayesian comparison among the constant, Burn, and modified Burn models yields 35 sources best described by the constant model, 8 by the Burn model, and 6 by the modified Burn model. The distribution of best-fit models across the full sample is shown in Figure~\ref{fig:FCBMB}e. Supplementary Figure~S2 presents the fits of these three models for each burst, and Supplementary Table~S3 provides the detailed Bayesian evidence comparison. Representative examples of each successfully fitted category (constant, Burn, modified Burn, and oscillation) are shown in Figure~\ref{fig:FCBMB}. Four bursts could not be adequately fitted by any of the four models; these are shown in Supplementary Figure~S3.

About half of the non-repeating bursts are best described by the constant model. This may be due to CHIME's limited bandwidth, which hinders the detection of linear polarization variations across its frequency range. Alternatively, the potential \(\sigma_{\mathrm{RM}}\) may be too small to produce significant changes within the CHIME band. For instance, for \(\sigma_{\mathrm{RM}} \lesssim 0.4\ \mathrm{rad\,m^{-2}}\), the polarization fraction drops from $\sim$99\% at \(800\ \mathrm{MHz}\) to $\sim$89\% at \(400\ \mathrm{MHz}\) (assuming \(100\%\) intrinsic linear polarization), a decrease of less than \(10\%\). Given that CHIME-detected FRBs typically have relatively low S/N and we adopt a sub-band S/N threshold of 5, our models are insensitive to non-repeating FRBs with \(\sigma_{\mathrm{RM}} \lesssim 0.4\ \mathrm{rad\,m^{-2}}\). Indeed, 14 (20\%) of these bursts are best described by the Burn or modified Burn model, with corresponding $\sigma_{\rm RM}$ values ranging from $0.5$ to $1.9\ \mathrm{rad\,m^{-2}}$. This absence of population-wide frequency-dependent depolarization is consistent with \citet{2024ApJ...968...50P} (hereafter P24), who similarly found no evidence for it. However, despite using the same underlying CHIME/FRB baseband data, P24 did not identify the same Burn-type depolarized FRBs that we report. This discrepancy likely arises from differing methods and thresholds. Specifically, P24 defined a depolarization ratio $f_{\rm depol} = (L/I)_{500} / (L/I)_{700}$, comparing the band-averaged linear polarization fraction between the 400–600 MHz and 600–800 MHz sub-bands of broadband FRBs, whereas we perform direct model fitting to the frequency dependence of $L/I$. The apparent discrepancy may also be exacerbated by the higher detection threshold adopted in P24; for instance, some FRBs below the $f_{\rm depol}=1$ dashed line in their Figure 11 could exhibit depolarization, particularly those approaching the dash-dotted curve predicted by the $\sigma_{\rm RM}$ model. Nevertheless, P24 did not classify any of these FRBs as depolarized. No cases with \(\sigma_{\mathrm{RM}} \leq 0.4\ \mathrm{rad\,m^{-2}}\) are found, likely due to the limitations of the data and our modeling approach. We note that none of these sources exhibit significant depolarization with \(\sigma_{\mathrm{RM}} \gtrsim 5\ \mathrm{rad\,m^{-2}}\).

\section{Conclusions}
\label{sec:con}
We have performed a systematic spectro-polarimetric analysis of 28 repeating and 70 non-repeating fast radio bursts detected by CHIME. For the repeating population, we measured the RM scatter parameter \(\sigma_{\mathrm{RM}}\) for each source, expanding the known sample from 14 to 36 sources (an increase by a factor of 2.6) compared to previous studies. The KDE of \(\sigma_{\mathrm{RM}}\) peaks at \(1.3\ \mathrm{rad\,m^{-2}}\), and approximately \(70\%\) of the repeaters exhibit \(\sigma_{\mathrm{RM}} \gtrsim 1\ \mathrm{rad\,m^{-2}}\), suggesting that the majority reside in complex magneto-ionic environments. However, we find no repeater with \(\sigma_{\mathrm{RM}} > 10\ \mathrm{rad\,m^{-2}}\); this absence is not a physical upper limit but rather a selection bias inherent to the CHIME band. At 800 MHz, a \(\sigma_{\mathrm{RM}}\) of \(7.6\ \mathrm{rad\,m^{-2}}\) already suppresses the linear polarization fraction to $\sim$10\%, below the typical detection threshold given the moderate S/N of the sample. Consequently, the most extreme magneto-ionic environments---if they exist---would remain largely invisible to current polarization-selected surveys, and the true distribution of \(\sigma_{\mathrm{RM}}\) among repeaters may be significantly broader than observed. There is also a secondary bump at \(\sigma_{\mathrm{RM}} \lesssim 0.5\ \mathrm{rad\,m^{-2}}\), which may indicate that some repeating FRBs reside in a less magneto-ionic environment. Combined with previously published results, especially from high-frequency observations, the broad range of \(\sigma_{\mathrm{RM}}\) values suggests that repeating FRBs may be at different evolutionary stages or may even have distinct physical origins.

For the non-repeating FRBs, we characterized their spectro-polarimetric properties by comparing four models: constant, Burn, modified Burn, and oscillation. No significant Burn-type depolarization with \(\sigma_{\mathrm{RM}} \gtrsim 5\ \mathrm{rad\,m^{-2}}\) is detected across the sample; in particular, half of the bursts are best described by a constant polarization fraction. Among the remaining sources, 8 favor the standard Burn model, 6 favor the modified Burn model, and 17 exhibit coherent oscillatory patterns in polarization fraction. Notably, four bursts could not be adequately fitted by any of the proposed models. For the 14 Burn-like FRBs, the \(\sigma_{\mathrm{RM}}\) values lie in the range \(0.5\) to \(1.9\ \mathrm{rad\,m^{-2}}\). No cases with \(\sigma_{\mathrm{RM}} \leq 0.4\ \mathrm{rad\,m^{-2}}\) are found, likely due to the limitations of the data and our modeling approach.

Wideband polarimetric observations, or the combination of polarimetric data from widely separated frequency bands, will enable a more comprehensive characterization of the spectro-polarimetric properties of FRBs. For repeating FRBs that are undetectable in polarization with CHIME, we anticipate that polarization measurements at higher radio frequencies will become feasible, potentially even leading to the discovery of new persistent radio sources associated with these repeaters. Meanwhile, for repeaters that show near-100\% linear polarization within the CHIME band, observations with Low Frequency Array (LOFAR) or Square Kilometre Array (SKA)-Low could provide precise constraints on \(\sigma_{\mathrm{RM}}\). Furthermore, wideband coverage will allow us to determine whether the polarization of non-repeating FRBs is indeed frequency-independent. The new facilities currently under construction are also poised to shape the field \citep{2019BAAS...51g.255H, burstt, 2024PASA...41..109L, 2026arXiv260413903C}. For example, DSA-2000 (0.7--2\,GHz) is expected to detect and localize FRBs at a rate of $\sim$$10^4$ yr$^{-1}$ \citep{2019BAAS...51g.255H}. Collectively, these efforts will significantly deepen our understanding of FRB environments and radiation physics.\vspace{6pt}

\supplementary{The following supporting information can be downloaded at
 \linksupplementary{s1}: Figure S1: Oscillation model fits for linear and circular polarization fractions vs.\ wavelength squared for each non-repeating FRB; Figure S2: Model fit results (Burn, modified Burn, and constant) for each non-repeating FRB; Figure S3: Observed linear polarization fraction vs.\ wavelength-squared for non-repeating FRBs that could not be fitted with the available models; Table S1: Rotation measure scatter ($\sigma_{\rm RM}$) for 28 repeating FRBs; Table S2: Oscillation model fitting parameters for non-repeating FRBs; Table S3: Bayesian model comparison results for non-repeating FRBs.}

\authorcontributions{Conceptualization, Y.F.; methodology, D.Z., and Y.F.; software, D.Z.; validation, D.Z. and Y.F.; formal analysis, D.Z.; investigation, D.Z. and Y.F.; resources, Y.F.; data curation, D.Z.; writing---original draft preparation, D.Z., Y.F., J.X. and J.F.; writing---review and editing, D.Z., Y.F., J.X., J.F. and C.X.; visualization, D.Z. and C.X.; supervision, Y.F.; project administration, Y.F.; funding acquisition, Y.F. All authors have read and agreed to the published version of the manuscript.}

\funding{This work is supported by National Key R\&D Program of China No. 2023YFB4503300, and the National Natural Science Foundation of China under grant 12522305.}
 
 \dataavailability{The original contributions presented in this study are included in the article/supplementary material. Further inquiries can be directed to the corresponding author.}

\acknowledgments{We thank the reviewers for their constructive reports.}

\conflictsofinterest{The authors declare no conflicts of interest.}

\printendnotes[custom]

\begin{adjustwidth}{-\extralength}{0cm}

\reftitle{References}

\PublishersNote{}
\end{adjustwidth}
\end{document}


\section{Supplementary Figures and Tables}

\begin{figure}[H]
    \centering
    \includegraphics[scale=0.21]{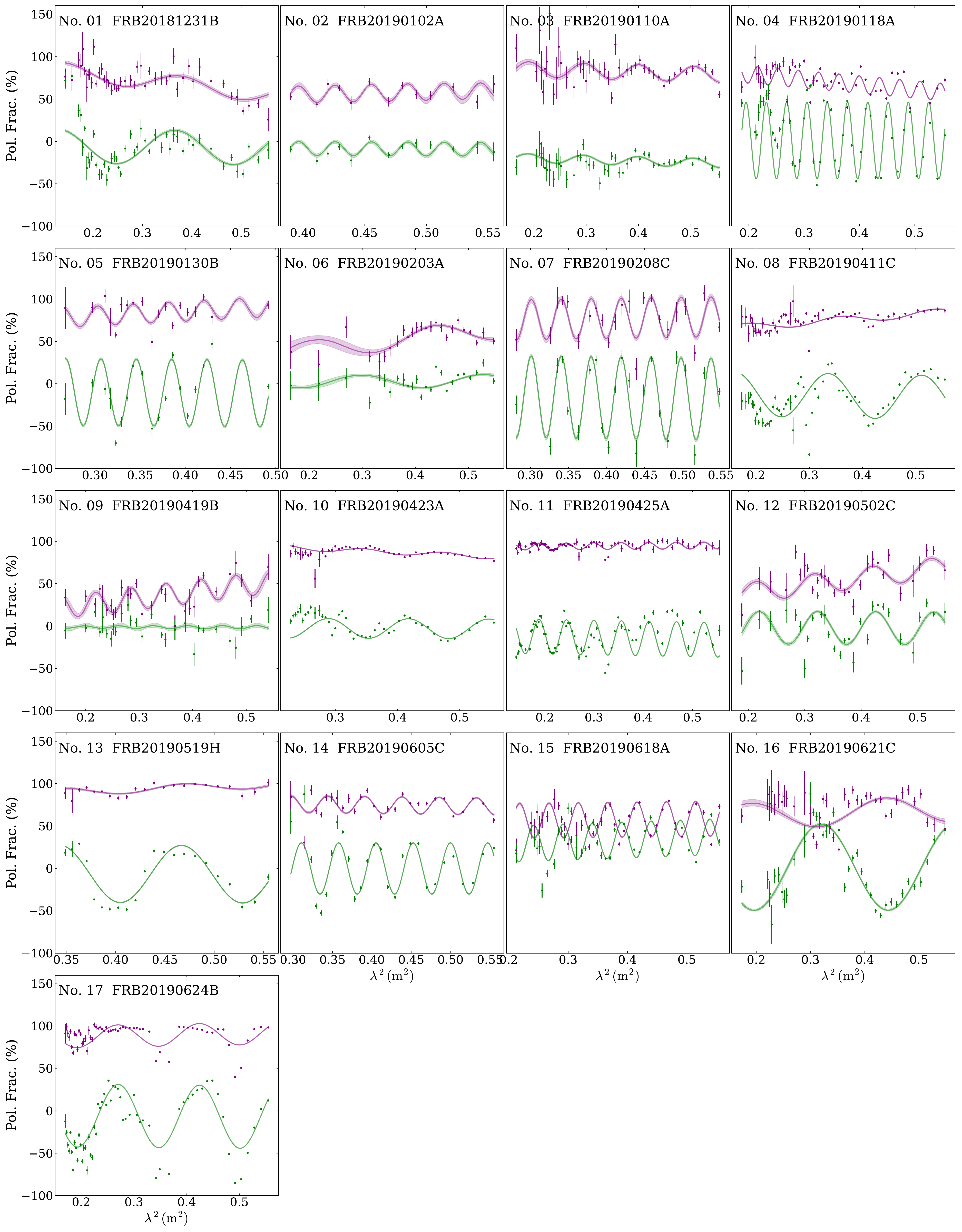}
    \caption{Oscillation model fits for linear and circular polarization fractions vs.\ wavelength squared for each non-repeating FRB. The purple and green data points represent the observed linear and circular polarization, with the model fits shown as shaded regions and solid lines.}
    \label{fig:fc_sup}
\end{figure}

\begin{figure}[H]
    \centering
    \includegraphics[scale=0.22]{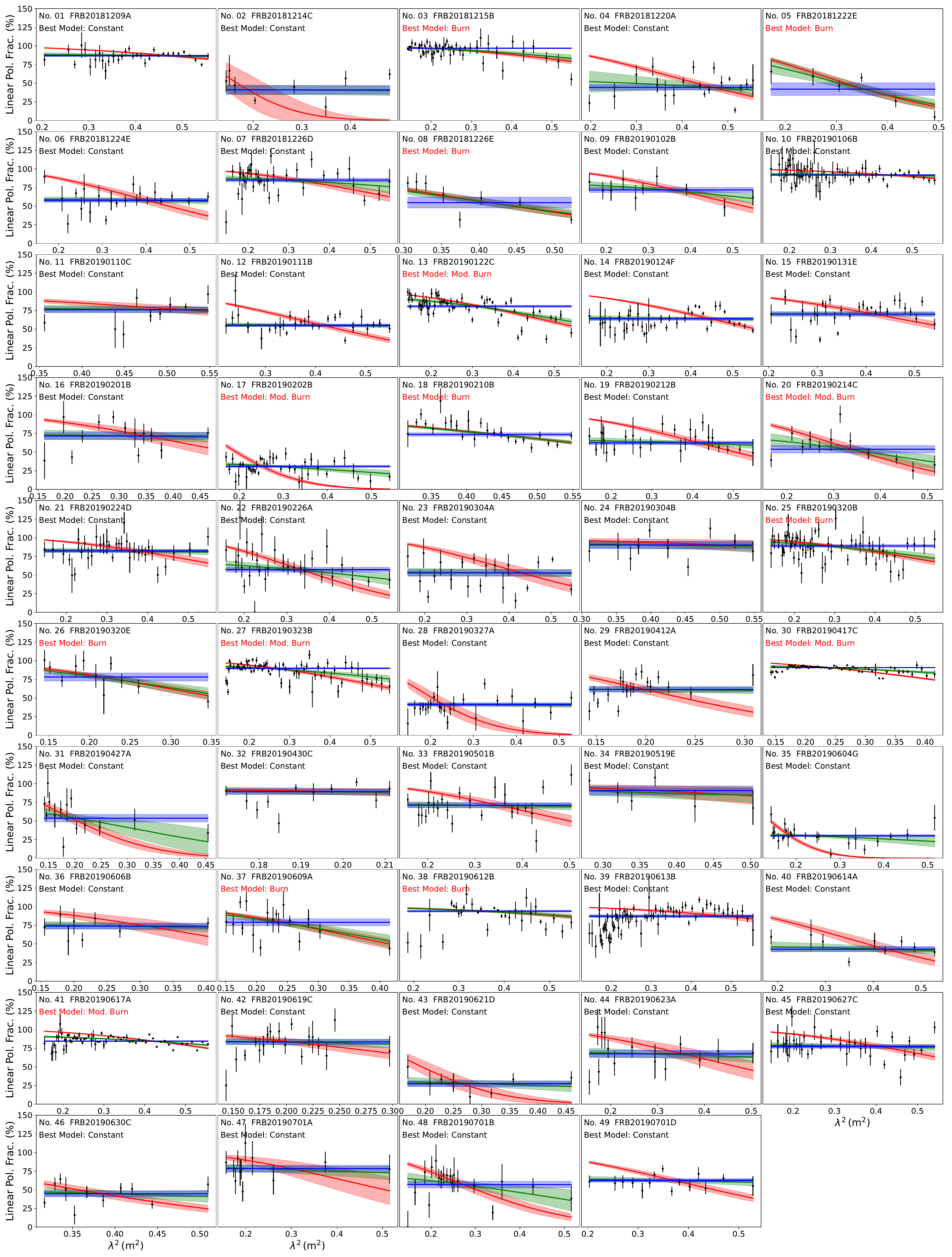}
    \caption{Model fit results for each non-repeating FRB. The data points (black) are fitted with three models: Burn (red), Mod.\ Burn (green), and Constant (blue). Each subplot represents a different FRB.}
    \label{fig:bmbc_sup}
\end{figure}

\begin{figure}[H]
    \centering
    \includegraphics[width=0.9\linewidth]{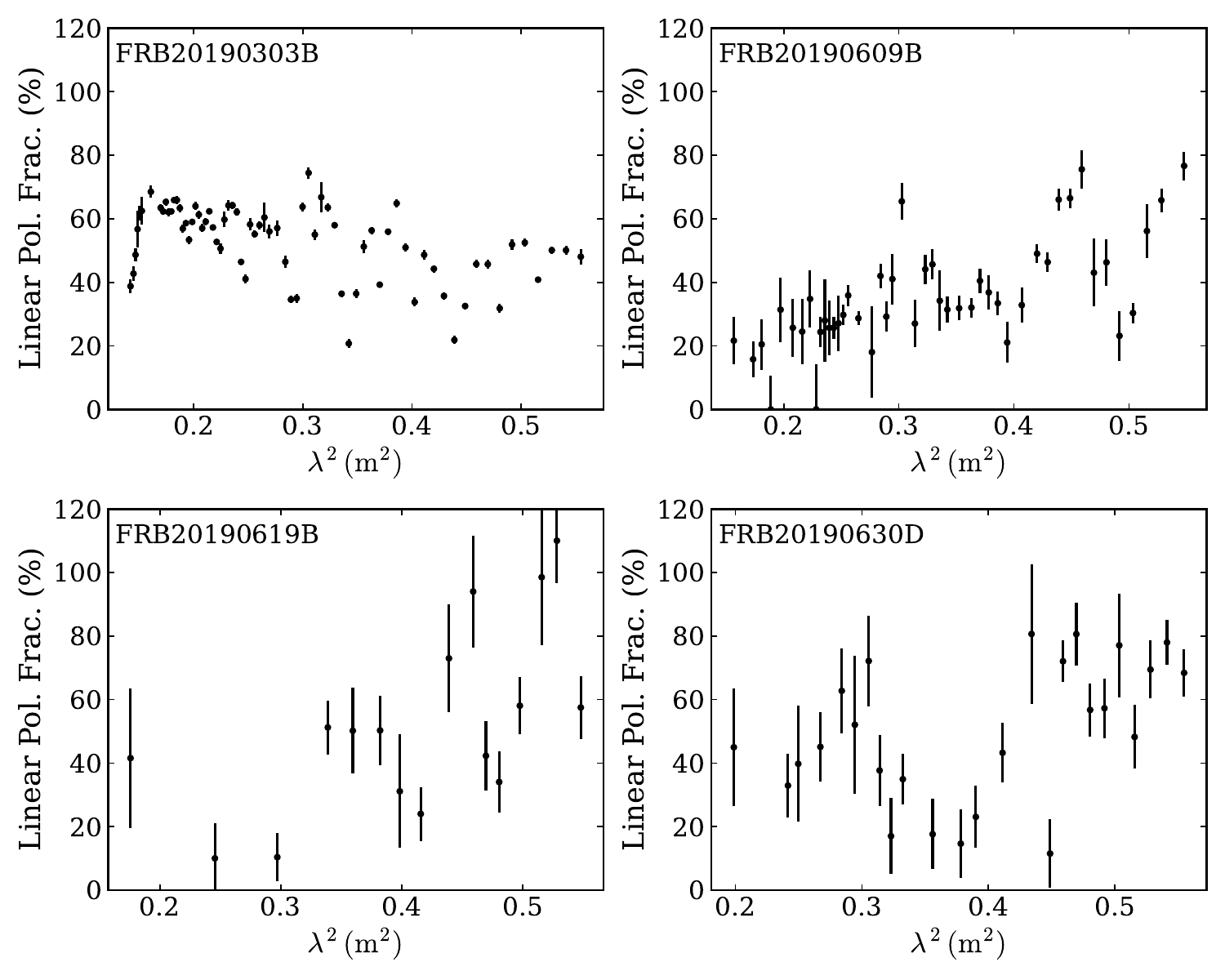}
    \caption{Observed linear polarization fraction vs.\ wavelength-squared for non-repeating FRBs that could not be fitted with the available models. Black points represent the data with associated errors.}
    \label{fig:other}
\end{figure}


\begin{center}
\setlength{\tabcolsep}{4pt}
\footnotesize
\setlength{\LTcapwidth}{\linewidth}
\begin{longtable}{c l l c c c c}
\caption{Rotation measure scatter ($\sigma_{\rm RM}$) for 28 repeating FRBs. Columns list the number (No.), FRB name, burst name, central observing frequency (MHz), measured linear polarization fraction ($L/I$), RM scatter $\sigma_{\rm RM}$ (rad m$^{-2}$), and the corresponding $\rm RM_{\rm FDF}$ (rad m$^{-2}$). All uncertainties are $1\sigma$.\label{tab:sigma_rm_repeaters}}\\
\toprule
No. & FRB Name & Burst Name & Cent.\ Freq.\ (MHz) & $L/I$ & $\sigma_{\rm RM}$ (rad m$^{-2}$) & $\rm RM_{\rm FDF}$ (rad m$^{-2}$)\\
\midrule
\endfirsthead
\toprule
No. & FRB Name & Burst Name & Cent.\ Freq.\ (MHz) & $L/I$ & $\sigma_{\rm RM}$ (rad m$^{-2}$) & $\rm RM_{\rm FDF}$ (rad m$^{-2}$) \\
\midrule
\endhead
\midrule
\multicolumn{7}{r}{{Continued on next page}} \\
\endfoot
\bottomrule
\endlastfoot
01 & FRB 20171019A & 20200801C & 421 & 0.851$\pm$0.015 & 0.56$\pm$0.11 & $-$2.0$\pm$0.3\\
02 & FRB 20180910A & 20200621D & 682 & 0.343$\pm$0.008 & 3.79$\pm$0.26 & $-$340.37$\pm$0.16\\
03 & FRB 20180916B & 20220619B & 423 & 0.945$\pm$0.006 & 0.33$\pm$0.16 & $-$53.30$\pm$0.12\\
03 & FRB 20180916B & 20221215G & 431 & 0.888$\pm$0.012 & 0.50$\pm$0.12 & $-$59.80$\pm$0.17\\
03 & FRB 20180916B & 20231108A & 487 & 0.716$\pm$0.019 & 1.08$\pm$0.12 & $-$53.8$\pm$0.3\\
03 & FRB 20180916B & 20230510G & 477 & 0.834$\pm$0.009 & 0.76$\pm$0.13 & $-$54.49$\pm$0.07\\
03 & FRB 20180916B & 20220312A & 685 & 0.64$\pm$0.02 & 2.47$\pm$0.23 & $-$54.1$\pm$2.3\\
03 & FRB 20180916B & 20220328A & 687 & 0.868$\pm$0.018 & 1.40$\pm$0.30 & $-$56.6$\pm$0.5\\
03 & FRB 20180916B & 20240520A & 457 & 0.968$\pm$0.006 & 0.30$\pm$0.24 & $-$56.14$\pm$0.06\\
03 & FRB 20180916B & 20240228A & 433 & 0.81$\pm$0.02 & 0.68$\pm$0.11 & $-$52.2$\pm$0.2\\
03 & FRB 20180916B & 20231123D & 435 & 0.74$\pm$0.03 & 0.82$\pm$0.11 & $-$51.8$\pm$0.4\\
03 & FRB 20180916B & 20231210E & 454 & 0.866$\pm$0.007 & 0.62$\pm$0.12 & $-$57.94$\pm$0.11\\
04 & FRB 20181119A & 20210608B & 701 & 0.274$\pm$0.014 & 4.40$\pm$0.32 & 355.2$\pm$1.1\\
05 & FRB 20190110C & 20190110C & 436 & 0.93$\pm$0.02 & 0.40$\pm$0.16 & 118.5$\pm$0.2\\
06 & FRB 20190117A & 20211114A & 501 & 0.476$\pm$0.019 & 1.70$\pm$0.13 & 28.9$\pm$0.2\\
07 & FRB 20190208A & 20230306D & 700 & 0.46$\pm$0.18 & 3.40$\pm$0.89 & 26.3$\pm$1.0\\
08 & FRB 20190303A & 20230913E & 682 & 0.805$\pm$0.007 & 1.70$\pm$0.25 & 371.7$\pm$0.3\\
08 & FRB 20190303A & 20231204A & 570 & 0.49$\pm$0.14 & 2.16$\pm$0.46 & 287.8$\pm$0.2\\
09 & FRB 20190430C & 20190430C & 664 & 0.913$\pm$0.019 & 1.05$\pm$0.34 & $-$71.25$\pm$0.17\\
10 & FRB 20190604A & 20210329A & 453 & 0.37$\pm$0.02 & 1.61$\pm$0.12 & $-$14.6$\pm$0.4\\
11 & FRB 20190609C & 20210113D & 450 & $<$ 0.55$\pm$0.18 & $>$ 1.23$\pm$0.35 & ---\\
11 & FRB 20190609C & 20201030B & 438 & 0.966$\pm$0.002 & 0.28$\pm$0.21 & $-$39.46$\pm$0.02\\
11 & FRB 20190609C & 20190609C & 444 & 0.9337$\pm$0.0015 & 0.41$\pm$0.16 & $-$3.7$\pm$0.3\\
12 & FRB 20190804E & 20200709C & 425 & 0.55$\pm$0.03 & 1.10$\pm$0.10 & $-$206.4$\pm$0.6\\
12 & FRB 20190804E & 20201228A & 534 & 0.809$\pm$0.015 & 1.03$\pm$0.16 & $-$196.0$\pm$0.2\\
12 & FRB 20190804E & 20201225B & 453 & 0.991$\pm$0.012 & 0.15$\pm$0.44 & $-$202.53$\pm$0.15\\
12 & FRB 20190804E & 20200629C & 482 & 0.435$\pm$0.002 & 1.67$\pm$0.12 & $-$200.75$\pm$0.14\\
13 & FRB 20190915D & 20200214B & 439 & 0.274$\pm$0.012 & 1.73$\pm$0.13 & $-$170.3$\pm$0.4\\
14 & FRB 20191013D & 20200515A & 585 & 0.357$\pm$0.013 & 2.73$\pm$0.19 & $-$35.7$\pm$0.2\\
15 & FRB 20191106C & 20211104B & 567 & 0.186$\pm$0.009 & 3.28$\pm$0.27 & $-$921.8$\pm$0.3\\
15 & FRB 20191106C & 20210212C & 450 & $<$ 0.33$\pm$0.15 & $>$ 1.68$\pm$0.36 & ---\\
15 & FRB 20191106C & 20201201A & 432 & 0.172$\pm$0.005 & 1.95$\pm$0.16 & $-$263.3$\pm$0.2\\
15 & FRB 20191106C & 20220118B & 540 & 0.219$\pm$0.004 & 2.83$\pm$0.21 & $-$1044.43$\pm$0.17\\
15 & FRB 20191106C & 20210617A & 456 & 0.209$\pm$0.002 & 2.05$\pm$0.16 & $-$468.5$\pm$0.2\\
16 & FRB 20200118D & 20200701A & 621 & 0.696$\pm$0.013 & 1.83$\pm$0.19 & 132.6$\pm$0.3\\
16 & FRB 20200118D & 20200118D & 446 & 0.258$\pm$0.002 & 1.82$\pm$0.13 & 125.4$\pm$0.8\\
17 & FRB 20200120E & 20231001A & 592 & 0.75$\pm$0.02 & 1.48$\pm$0.18 & $-$24.45$\pm$0.13\\
17 & FRB 20200120E & 20210423G & 678 & 0.993$\pm$0.007 & 0.30$\pm$1.10 & $-$28.78$\pm$0.14\\
17 & FRB 20200120E & 20200120E & 692 & 0.95$\pm$0.02 & 0.85$\pm$0.47 & $-$29.8$\pm$0.2\\
18 & FRB 20200127B & 20200127B & 562 & 0.703$\pm$0.010 & 1.48$\pm$0.15 & 39.32$\pm$0.05\\
18 & FRB 20200127B & 20200219B & 586 & 0.909$\pm$0.007 & 0.83$\pm$0.24 & 40.58$\pm$0.04\\
19 & FRB 20200202A & 20201014B & 750 & 0.833$\pm$0.009 & 1.89$\pm$0.32 & 52.6$\pm$0.2\\
19 & FRB 20200202A & 20230604B & 674 & 0.23$\pm$0.09 & 4.33$\pm$0.66 & 61.5$\pm$1.0\\
20 & FRB 20200223B & 20200702C & 697 & 0.182$\pm$0.002 & 4.99$\pm$0.40 & 70.4$\pm$0.4\\
20 & FRB 20200223B & 20210115C & 650 & $<$ 0.328$\pm$0.008 & $>$ 3.51$\pm$0.24 & ---\\
21 & FRB 20200619A & 20201022D & 550 & $<$ 0.31$\pm$0.15 & $>$ 2.58$\pm$0.56 & ---\\
21 & FRB 20200619A & 20210130E & 450 & $<$ 0.21$\pm$0.10 & $>$ 1.99$\pm$0.34 & ---\\
22 & FRB 20200809E & 20200809E & 430 & 0.7$\pm$0.2 & 0.87$\pm$0.36 & $-$40.6$\pm$0.6\\
22 & FRB 20200809E & 20201018C & 519 & 0.8$\pm$0.2 & 1.00$\pm$0.58 & $-$36.5$\pm$0.3\\
22 & FRB 20200809E & 20210208B & 541 & 0.58$\pm$0.03 & 1.70$\pm$0.16 & $-$36.3$\pm$0.5\\
23 & FRB 20200926A & 20201223A & 450 & $<$ 0.51$\pm$0.16 & $>$ 1.31$\pm$0.32 & ---\\
23 & FRB 20200926A & 20230725D & 620 & 0.266$\pm$0.007 & 3.48$\pm$0.25 & 274.1$\pm$0.4\\
24 & FRB 20200929C & 20220209A & 441 & 0.908$\pm$0.011 & 0.48$\pm$0.14 & $-$12.25$\pm$0.11\\
24 & FRB 20200929C & 20210930A & 487 & 0.531$\pm$0.008 & 1.48$\pm$0.11 & $-$42.37$\pm$0.13\\
24 & FRB 20200929C & 20210326B & 475 & 0.874$\pm$0.007 & 0.65$\pm$0.14 & 11.34$\pm$0.07\\
24 & FRB 20200929C & 20210314A & 450 & 0.897$\pm$0.016 & 0.53$\pm$0.14 & $-$1.55$\pm$0.19\\
24 & FRB 20200929C & 20210313B & 539 & 0.906$\pm$0.007 & 0.72$\pm$0.20 & 9.23$\pm$0.11\\
24 & FRB 20200929C & 20201203C & 445 & 0.65$\pm$0.02 & 1.02$\pm$0.10 & $-$8.4$\pm$0.4\\
24 & FRB 20200929C & 20201125B & 608 & 0.879$\pm$0.008 & 1.04$\pm$0.23 & $-$4.85$\pm$0.09\\
25 & FRB 20201114A & 20201219A & 568 & 0.541$\pm$0.012 & 1.99$\pm$0.15 & 1348.7$\pm$0.3\\
26 & FRB 20201124A & 20210331A & 697 & 0.439$\pm$0.009 & 3.47$\pm$0.24 & $-$576.3$\pm$0.3\\
26 & FRB 20201124A & 20210526D & 451 & 0.266$\pm$0.002 & 1.84$\pm$0.13 & $-$602.01$\pm$0.05\\
26 & FRB 20201124A & 20210327A & 681 & 0.263$\pm$0.009 & 4.22$\pm$0.30 & $-$543.6$\pm$0.3\\
27 & FRB 20201130A & 20210117E & 450 & 0.46$\pm$0.020 & 1.40$\pm$0.11 & 188.9$\pm$0.6\\
27 & FRB 20201130A & 20210118B & 602 & 0.633$\pm$0.011 & 1.93$\pm$0.17 & 184.01$\pm$0.13\\
27 & FRB 20201130A & 20210327F & 623 & 0.43$\pm$0.11 & 2.81$\pm$0.47 & 182.94$\pm$0.16\\
27 & FRB 20201130A & 20210114B & 622 & 0.618$\pm$0.013 & 2.11$\pm$0.18 & 183.53$\pm$0.15\\
27 & FRB 20201130A & 20201225D & 559 & 0.475$\pm$0.013 & 2.12$\pm$0.15 & 183.62$\pm$0.16\\
28 & FRB 20201221B & 20210303F & 510 & $<$ 0.155$\pm$0.008 & $>$ 2.79$\pm$0.24 & ---\\
28 & FRB 20201221B & 20210302E & 451 & 0.455$\pm$0.013 & 1.42$\pm$0.10 & $-$1.7$\pm$0.2\\
28 & FRB 20201221B & 20210224A & 425 & $<$ 0.6$\pm$0.2 & $>$ 1.02$\pm$0.34 & ---\\
\end{longtable}
\end{center}


\begin{center}
\setlength{\tabcolsep}{4pt}
\footnotesize
\setlength{\LTcapwidth}{\linewidth}
\begin{longtable}{c c c l l c c c c c }
\caption{Fitting results for the non-repeating FRBs using the oscillation model. The fractional linear polarization ($L/I$) and circular polarization ($V/I$) are fitted as functions of $\lambda^2$. Columns list the FRB number (No.), FRB name, signal-to-noise ratio, polarization type, best-fit parameters $A$, $\omega$, $\phi_0$, $k$, and $c$, along with their $1\sigma$ uncertainties.\label{tab:fc_params}}\\
\toprule
No. & FRB Name & $\rm S/N$ & Pol. & $A$ & $\omega$ & $\phi_{0}$ & $k$ & $c$ & $\rm RM_{\rm QU}$ (rad m$^{-2}$)\\
\midrule
\endfirsthead
\toprule
No. & FRB Name & $\rm S/N$ & Pol. & $A$ & $\omega$ & $\phi_{0}$ & $k$ & $c$ & $\rm RM_{\rm QU}$ (rad m$^{-2}$)\\
\midrule
\endhead
\midrule
\multicolumn{10}{r}{{Continued on next page}} \\
\endfoot
\bottomrule
\endlastfoot
01 & FRB 20181231B & 48.0 &$L/I$ & 10.1$\pm$1.4 & 26.52$\pm$0.50 & $-$2.16$\pm$0.22 & $-$66.7$\pm$9.2 & 92.4$\pm$3.1 & 10.51$\pm$0.01 \\
 &  &  &$V/I$ & 20.1$\pm$1.1 & 26.52$\pm$0.50 & $-$1.85$\pm$0.14 & $-$5.09$\pm$0.63 & $-$5.09$\pm$0.63 & \\
02 & FRB 20190102A & 21.2 &$L/I$ & 10.3$\pm$2.1 & 215$\pm$26 & $-$1.2$\pm$1.1 & 25$\pm$37 & 46$\pm$17 & $-$198.40$\pm$0.04 \\
 &  &  &$V/I$ & 8.5$\pm$1.7 & 215$\pm$26 & $-$1.2$\pm$1.0 & $-$6.06$\pm$0.74 & $-$6.06$\pm$0.74 & \\
03 & FRB 20190110A & 39.0 &$L/I$ & 9.2$\pm$1.3 & 59.7$\pm$1.1 & 1.8$\pm$1.9 & $-$18$\pm$11 & 88.5$\pm$4.8 & $-$30.74$\pm$0.06 \\
 &  &  &$V/I$ & 5.29$\pm$0.89 & 59.7$\pm$1.1 & 1.9$\pm$1.9 & $-$16.63$\pm$0.46 & $-$16.63$\pm$0.46 & \\
04 & FRB 20190118A & 108.6 &$L/I$ & 10.24$\pm$0.39 & 171.06$\pm$0.09 & 2.30$\pm$0.05 & $-$59.0$\pm$3.5 & 90.7$\pm$1.4 & $-$0.670$\pm$0.001 \\
 &  &  &$V/I$ & 45.14$\pm$0.33 & 171.06$\pm$0.09 & $-$0.36$\pm$0.04 & 0.74$\pm$0.17 & 0.74$\pm$0.17 & \\
05 & FRB 20190130B & 22.7 &$L/I$ & 11.7$\pm$1.7 & 160.61$\pm$0.52 & $-$1.6$\pm$2.5 & 55$\pm$24 & 63.4$\pm$9.3 & 77.54$\pm$0.03 \\
 &  &  &$V/I$ & 39.4$\pm$1.4 & 160.61$\pm$0.52 & 2.60$\pm$0.20 & $-$7.72$\pm$0.64 & $-$7.72$\pm$0.64 & \\
06 & FRB 20190203A & 31.8 &$L/I$ & 12.3$\pm$1.6 & 27.2$\pm$2.2 & 1.5$\pm$1.6 & 68$\pm$19 & 26.8$\pm$8.4 & $-$337.8$\pm$0.1 \\
 &  &  &$V/I$ & 7.9$\pm$1.0 & 27.2$\pm$2.2 & $-$0.2$\pm$1.0 & 1.98$\pm$0.59 & 1.98$\pm$0.59 & \\
07 & FRB 20190208C & 34.8 &$L/I$ & 23.7$\pm$2.3 & 159.79$\pm$0.60 & $-$2.3$\pm$1.1 & 13$\pm$25 & 72$\pm$11 & $-$80.70$\pm$0.10 \\
 &  &  &$V/I$ & 48.3$\pm$2.0 & 159.79$\pm$0.60 & $-$2.3$\pm$1.1 & $-$11.83$\pm$0.96 & $-$11.83$\pm$0.96 & \\
08 & FRB 20190411C & 217.9 &$L/I$ & 4.27$\pm$0.31 & 35.79$\pm$0.11 & 1.33$\pm$0.08 & 47.3$\pm$2.6 & 58.37$\pm$0.97 & $-$20.42$\pm$0.02 \\
 &  &  &$V/I$ & 25.98$\pm$0.22 & 35.79$\pm$0.11 & 2.13$\pm$0.04 & $-$10.53$\pm$0.12 & $-$10.53$\pm$0.12 & \\
09 & FRB 20190419B & 44.8 &$L/I$ & 13.7$\pm$1.8 & 96.2$\pm$2.3 & $-$0.63$\pm$0.84 & 75$\pm$14 & 10.2$\pm$4.8 & $-$10.63$\pm$0.08 \\
 &  &  &$V/I$ & 2.8$\pm$2.0 & 96.2$\pm$2.3 & 0.4$\pm$1.7 & $-$0.97$\pm$0.85 & $-$0.97$\pm$0.85 & \\
10 & FRB 20190423A & 113.1 &$L/I$ & 2.72$\pm$0.17 & 48.69$\pm$0.15 & $-$2.60$\pm$0.09 & $-$32.39$\pm$0.48 & 99.82$\pm$0.17 & $-$22.933$\pm$0.007 \\
 &  &  &$V/I$ & 11.62$\pm$0.11 & 48.69$\pm$0.15 & 0.10$\pm$0.07 & $-$2.14$\pm$0.06 & $-$2.14$\pm$0.06 & \\
11 & FRB 20190425A & 192.8 &$L/I$ & 3.95$\pm$0.32 & 112.38$\pm$0.12 & $-$0.65$\pm$0.07 & 3.4$\pm$2.7 & 93.30$\pm$0.72 & 57.043$\pm$0.002 \\
 &  &  &$V/I$ & 20.38$\pm$0.21 & 112.38$\pm$0.12 & $-$0.73$\pm$0.03 & $-$10.48$\pm$0.10 & $-$10.48$\pm$0.10 & \\
12 & FRB 20190502C & 34.7 &$L/I$ & 12.3$\pm$1.7 & 61.0$\pm$1.0 & 1.11$\pm$0.45 & 87.4$\pm$9.9 & 21.9$\pm$3.9 & 36.5$\pm$0.1 \\
 &  &  &$V/I$ & 19.4$\pm$1.3 & 61.0$\pm$1.0 & 0.76$\pm$0.41 & $-$1.67$\pm$0.70 & $-$1.67$\pm$0.70 & \\
13 & FRB 20190519H & 68.3 &$L/I$ & 4.51$\pm$0.72 & 50.63$\pm$0.12 & 2.3$\pm$1.9 & 42.9$\pm$9.0 & 75.2$\pm$4.0 & $-$25.03$\pm$0.03 \\
 &  &  &$V/I$ & 33.83$\pm$0.42 & 50.63$\pm$0.12 & 3.07$\pm$0.05 & $-$4.58$\pm$0.22 & $-$4.58$\pm$0.22 & \\
14 & FRB 20190605C & 76.5 &$L/I$ & 10.01$\pm$0.68 & 132.86$\pm$0.23 & $-$0.10$\pm$0.13 & $-$11.3$\pm$7.2 & 79.3$\pm$3.4 & $-$0.1$\pm$0.04 \\
 &  &  &$V/I$ & 30.19$\pm$0.51 & 132.86$\pm$0.23 & $-$1.98$\pm$0.11 & $-$0.09$\pm$0.24 & $-$0.09$\pm$0.24 & \\
15 & FRB 20190618A & 81.4 &$L/I$ & 20.53$\pm$0.58 & 127.07$\pm$0.29 & $-$1.03$\pm$0.14 & 7.9$\pm$6.1 & 54.4$\pm$2.7 & 62.958$\pm$0.001 \\
 &  &  &$V/I$ & 21.15$\pm$0.51 & 127.07$\pm$0.29 & 2.25$\pm$0.13 & 23.98$\pm$0.26 & 23.98$\pm$0.26 & \\
16 & FRB 20190621C & 60.5 &$L/I$ & 15.6$\pm$1.4 & 25.39$\pm$0.29 & 1.7$\pm$2.4 & 24$\pm$14 & 57.1$\pm$5.7 & $-$2.06$\pm$0.07 \\
 &  &  &$V/I$ & 50.8$\pm$1.1 & 25.39$\pm$0.29 & $-$0.26$\pm$0.12 & 1.08$\pm$0.61 & 1.08$\pm$0.61 & \\
17 & FRB 20190624B & 968.5 &$L/I$ & 13.01$\pm$0.08 & 40.78$\pm$0.01 & 3.14$\pm$0.00 & 10.14$\pm$0.59 & 85.57$\pm$0.26 & 4.00503$\pm$0.00006 \\
 &  &  &$V/I$ & 37.08$\pm$0.06 & 40.78$\pm$0.01 & 3.12$\pm$0.01 & $-$4.81$\pm$0.03 & $-$4.81$\pm$0.03 & \\
\end{longtable}
\end{center}


\begin{center}
\setlength{\tabcolsep}{3pt}
\footnotesize
\setlength{\LTcapwidth}{\linewidth}
\begin{longtable}{c c c c c c c c c c c}
\caption{Bayesian model comparison for non-repeating FRBs. Columns list the FRB number (No.), FRB name, total intensity signal-to-noise ratio (S/N), differences in $\log_{10}$ Bayesian evidence relative to the best model for the Burn ($\Delta \log_{10} E_{\rm b}$), modified Burn ($\Delta \log_{10} E_{\rm mb}$), and constant ($\Delta \log_{10} E_{\rm c}$) models, the preferred model (Best Model), RM scatter $\sigma_{\rm RM}$, modified RM scatter $\sigma'_{\rm RM}$ (rad m$^{-2}$), and intrinsic linear polarization fraction $P_0$. The model with $\Delta \log_{10} E = 0$ is the preferred one. Uncertainties are $1\sigma$.\label{tab:bayesian_comparison}}\\
\toprule
No. & FRB Name & $\rm S/N$ & $\Delta \log_{10}{E_{\rm b}}$ & $\Delta \log_{10}{E_{\rm mb}}$ & $\Delta \log_{10}{E_{\rm c}}$ & Best Model & $\sigma_{\rm RM}$ (rad m$^{-2}$) & $\sigma^{\prime}_{\rm RM}$ & $P_0$ & $\rm RM_{\rm QU}$ (rad m$^{-2}$)\\
\midrule
\endfirsthead
\toprule
No. & FRB Name & $\rm S/N$ & $\Delta \log_{10}{E_{\rm b}}$ & $\Delta \log_{10}{E_{\rm mb}}$ & $\Delta \log_{10}{E_{\rm c}}$ & Best Model & $\sigma_{\rm RM}$ (rad m$^{-2}$) & $\sigma^{\prime}_{\rm RM}$ & $P_0$ & $\rm RM_{\rm QU}$ (rad m$^{-2}$)\\
\midrule
\endhead
\midrule
\multicolumn{11}{r}{{Continued on next page}} \\
\endfoot
\bottomrule
\endlastfoot
01 & FRB 20181209A & 38.1 & 4.74 & 2.17 & 0.00 & Const. & --- & --- & 0.87$\pm$0.01 & 106.81$\pm$0.01 \\
02 & FRB 20181214C & 11.7 & 20.51 & 2.46 & 0.00 & Const. & --- & --- & 0.41$\pm$0.06 & 23.8$\pm$0.2 \\
03 & FRB 20181215B & 125.9 & 0.00 & 1.30 & 1.72 & Burn & 0.62$\pm$0.05 & --- & --- & 8.18$\pm$0.03 \\
04 & FRB 20181220A & 39.0 & 7.74 & 1.38 & 0.00 & Const. & --- & --- & 0.44$\pm$0.04 & $-$0.64$\pm$0.07 \\
05 & FRB 20181222E & 8.4 & 0.00 & 0.66 & 3.23 & Burn & 1.86$\pm$0.13 & --- & --- & $-$92.9$\pm$0.1 \\
06 & FRB 20181224E & 21.5 & 17.99 & 2.44 & 0.00 & Const. & --- & --- & 0.58$\pm$0.03 & 0.67$\pm$0.09 \\
07 & FRB 20181226D & 34.9 & 4.50 & 1.47 & 0.00 & Const. & --- & --- & 0.85$\pm$0.02 & 64.73$\pm$0.05 \\
08 & FRB 20181226E & 15.7 & 0.00 & 1.20 & 4.89 & Burn & 1.32$\pm$0.08 & --- & --- & $-$0.2$\pm$0.2 \\
09 & FRB 20190102B & 13.6 & 2.22 & 1.46 & 0.00 & Const. & --- & --- & 0.72$\pm$0.04 & $-$13.31$\pm$0.03 \\
10 & FRB 20190106B & 64.5 & 4.23 & 2.35 & 0.00 & Const. & --- & --- & 0.92$\pm$0.01 & 72.99$\pm$0.03 \\
11 & FRB 20190110C & 15.1 & 3.23 & 2.43 & 0.00 & Const. & --- & --- & 0.76$\pm$0.04 & 118.4$\pm$0.3 \\
12 & FRB 20190111B & 48.7 & 28.32 & 2.36 & 0.00 & Const. & --- & --- & 0.55$\pm$0.02 & 331.9$\pm$0.1 \\
13 & FRB 20190122C & 47.0 & 7.34 & 0.00 & 35.91 & Mod.\ Burn & --- & 0.86$\pm$0.07 & 0.93$\pm$0.03 & 50.71$\pm$0.02 \\
14 & FRB 20190124F & 65.6 & 53.08 & 2.40 & 0.00 & Const. & --- & --- & 0.64$\pm$0.02 & 4.62$\pm$0.09 \\
15 & FRB 20190131E & 46.3 & 35.84 & 2.82 & 0.00 & Const. & --- & --- & 0.70$\pm$0.03 & $-$0.56$\pm$0.01 \\
16 & FRB 20190201B & 17.1 & 5.44 & 2.24 & 0.00 & Const. & --- & --- & 0.72$\pm$0.05 & $-$156.62$\pm$0.09 \\
17 & FRB 20190202B & 45.9 & 50.10 & 0.00 & 0.41 & Mod.\ Burn & --- & 0.94$\pm$0.26 & 0.36$\pm$0.03 & 571.3$\pm$0.2 \\
18 & FRB 20190210B & 27.0 & 0.00 & 1.74 & 9.24 & Burn & 0.88$\pm$0.03 & --- & --- & $-$359.42$\pm$0.09 \\
19 & FRB 20190212B & 23.4 & 13.40 & 1.99 & 0.00 & Const. & --- & --- & 0.62$\pm$0.02 & $-$175.7$\pm$0.2 \\
20 & FRB 20190214C & 22.9 & 2.87 & 0.00 & 0.24 & Mod.\ Burn & --- & 1.10$\pm$0.34 & 0.71$\pm$0.11 & $-$1169.6$\pm$0.1 \\
21 & FRB 20190224D & 40.1 & 6.51 & 2.42 & 0.00 & Const. & --- & --- & 0.82$\pm$0.02 & $-$41.86$\pm$0.09 \\
22 & FRB 20190226A & 35.0 & 8.68 & 0.76 & 0.00 & Const. & --- & --- & 0.57$\pm$0.03 & $-$234.1$\pm$0.1 \\
23 & FRB 20190304A & 23.3 & 23.91 & 2.47 & 0.00 & Const. & --- & --- & 0.53$\pm$0.04 & 78.0$\pm$0.1 \\
24 & FRB 20190304B & 16.2 & 1.63 & 2.25 & 0.00 & Const. & --- & --- & 0.90$\pm$0.05 & $-$39$\pm$0.1 \\
25 & FRB 20190320B & 70.6 & 0.00 & 0.32 & 2.25 & Burn & 0.80$\pm$0.06 & --- & --- & 53.36$\pm$0.06 \\
26 & FRB 20190320E & 16.5 & 0.00 & 1.17 & 2.11 & Burn & 1.63$\pm$0.14 & --- & --- & 75.74$\pm$0.01 \\
27 & FRB 20190323B & 87.8 & 11.05 & 0.00 & 5.50 & Mod.\ Burn & --- & 0.60$\pm$0.10 & 0.94$\pm$0.02 & 229.058$\pm$0.003 \\
28 & FRB 20190327A & 24.1 & 41.36 & 2.42 & 0.00 & Const. & --- & --- & 0.41$\pm$0.02 & 11.2$\pm$0.1 \\
29 & FRB 20190412A & 16.5 & 12.40 & 2.23 & 0.00 & Const. & --- & --- & 0.62$\pm$0.04 & 167.58$\pm$0.07 \\
30 & FRB 20190417C & 605.6 & 90.37 & 0.00 & 34.41 & Mod.\ Burn & --- & 0.59$\pm$0.07 & 0.94$\pm$0.01 & $-$475.4005$\pm$0.0002 \\
31 & FRB 20190427A & 14.1 & 1.26 & 0.30 & 0.00 & Const. & --- & --- & 0.54$\pm$0.05 & $-$524.1$\pm$0.3 \\
32 & FRB 20190430C & 16.4 & 1.70 & 2.06 & 0.00 & Const. & --- & --- & 0.89$\pm$0.04 & $-$70.4$\pm$0.5 \\
33 & FRB 20190501B & 22.8 & 11.34 & 2.35 & 0.00 & Const. & --- & --- & 0.71$\pm$0.03 & 121.35$\pm$0.09 \\
34 & FRB 20190519E & 14.6 & 1.16 & 2.06 & 0.00 & Const. & --- & --- & 0.90$\pm$0.06 & $-$17$\pm$0.4 \\
35 & FRB 20190604G & 23.9 & 21.55 & 1.50 & 0.00 & Const. & --- & --- & 0.30$\pm$0.01 & 364.6$\pm$0.4 \\
36 & FRB 20190606B & 16.1 & 4.77 & 2.15 & 0.00 & Const. & --- & --- & 0.74$\pm$0.04 & 16.7$\pm$0.2 \\
37 & FRB 20190609A & 16.8 & 0.00 & 1.04 & 2.34 & Burn & 1.42$\pm$0.14 & --- & --- & $-$44.2$\pm$0.2 \\
38 & FRB 20190612B & 43.5 & 0.00 & 2.05 & 3.25 & Burn & 0.49$\pm$0.04 & --- & --- & $-$0.88$\pm$0.08 \\
39 & FRB 20190613B & 112.7 & 42.74 & 3.15 & 0.00 & Const. & --- & --- & 0.87$\pm$0.02 & $-$22.36$\pm$0.03 \\
40 & FRB 20190614A & 16.4 & 8.53 & 1.92 & 0.00 & Const. & --- & --- & 0.43$\pm$0.03 & $-$1.7$\pm$0.2 \\
41 & FRB 20190617A & 128.5 & 109.87 & 0.00 & 96.57 & Mod.\ Burn & --- & 0.49$\pm$0.04 & 0.92$\pm$0.01 & 94.068$\pm$0.001 \\
42 & FRB 20190619C & 21.5 & 2.91 & 2.13 & 0.00 & Const. & --- & --- & 0.83$\pm$0.04 & 226.9$\pm$0.3 \\
43 & FRB 20190621D & 12.5 & 4.69 & 1.85 & 0.00 & Const. & --- & --- & 0.27$\pm$0.04 & $-$759.2$\pm$0.2 \\
44 & FRB 20190623A & 17.4 & 5.21 & 2.06 & 0.00 & Const. & --- & --- & 0.67$\pm$0.05 & $-$161.54$\pm$0.07 \\
45 & FRB 20190627C & 30.0 & 9.63 & 2.19 & 0.00 & Const. & --- & --- & 0.78$\pm$0.02 & 68.47$\pm$0.06 \\
46 & FRB 20190630C & 19.7 & 4.14 & 1.85 & 0.00 & Const. & --- & --- & 0.45$\pm$0.04 & 641.7$\pm$0.2 \\
47 & FRB 20190701A & 18.9 & 3.56 & 2.07 & 0.00 & Const. & --- & --- & 0.78$\pm$0.05 & 154.3$\pm$0.2 \\
48 & FRB 20190701B & 17.3 & 2.95 & 1.01 & 0.00 & Const. & --- & --- & 0.57$\pm$0.04 & $-$533.5$\pm$0.5 \\
49 & FRB 20190701D & 16.6 & 8.84 & 2.15 & 0.00 & Const. & --- & --- & 0.63$\pm$0.02 & 137.01$\pm$0.04 \\
\end{longtable}
\end{center}